\documentclass[aps,twocolumn,a4paper,10pt]{revtex4-1}
\usepackage{mathrsfs}
\usepackage{graphicx}
\usepackage{latexsym}
\usepackage{amsmath}
\usepackage{amssymb}
\usepackage{textcomp}
\usepackage{amsbsy}
\usepackage{epstopdf}

\begin{document}
\title{\large \bf ``Brane localized energy density'' stabilizes the modulus in higher dimensional warped spacetime}
\author{Tanmoy Paul}
\email{pul.tnmy9@gmail.com}
\affiliation{Department of Theoretical Physics,\\
Indian Association for the Cultivation of Science,\\
2A $\&$ 2B Raja S.C. Mullick Road,\\
Kolkata - 700 032, India.\\}

\begin{abstract}
We consider a five dimensional AdS spacetime with two 3-brane scenario where the 
hidden brane contains a certain amount of positive constant energy density. 
In this model, we examine the possibility of modulus stabilization. 
Our result reveals that the non-zero value of ``hidden brane energy density'' is 
sufficient to stabilize the two brane system. Moreover we 
scan the parametric space for which the modulus (or radion) is going to be stabilized 
without sacrificing the conditions necessary to solve the gauge hierarchy problem.
Finally we obtain the mass and coupling parameters 
of radion field in this higher dimensional braneworld scenario.
\end{abstract}
\maketitle

\section{Introduction}
The possible interactions between fundamental particles are best described by 
Standard Model (SM) of particle physics. Despite its enormous successes, the model 
suffers with a divergence due to the radiative correction of Higgs mass which 
may run up to Planck scale. An unnatural fine tuning is needed to confine the 
Higgs mass within TeV scale.

Many attempts have been made to solve this problem and the theory of extra dimensional 
models \cite{arkani,horava,RS1,kaloper,cohen,burgess,chodos} are one of them. 
Extra dimensional models also have a natural outcome from String theory. 
Depending on their geometry, these models are compactified under various compactification schemes. 
Our usual four dimensional universe which is considered to be a 3-brane embedded within higher dimensional
spacetime, emerges as a low 
energy effective theory \cite{kanno,shiromizu} and contains signatures of compactified extra dimension.

Among all the extra dimensional models proposed so far, Randall-Sundrum (RS) warped 
model \cite{RS1} earned a special attention since it resolves the gauge hierarchy problem without 
introducing any intermediate scale (between TeV and Planck scale) in the theory. In RS 
model, the full spaetime is of five dimensional AdS nature where the extra dimension is spacelike 
and $S^1/Z_2$ orbilfolded. The orbifolded fixed points are identified with two 3-branes 
embedded in the five dimensional spacetime. The separation between the branes is assumed to be 
$\sim$ Planck length so that the hierarchy problem can be solved. 

However, without any stabilization 
mechanism, two brane system can collapse due to the intervening gravity. So, like other 
higher dimensional braneworld scenario, one of the crucial aspects of RS model is 
to stabilize the inter-brane separation (known as modulus 
or radion). For this purpose, one needs to generate a radion potential with a stable minimum. 
Goldberger and Wise (GW) proposed a useful mechanism \cite{GW1} to construct such a radion potential by imposing 
a massive scalar field in the bulk where the scalar field has appropriate values on the branes. 
In GW mechanism, it is not required to introduce any hierarchy between the boundary values 
of the scalar field for stabilizing the modulus at a value consistent with that proposed in RS model in order to solve 
the gauge hierarchy problem. Subsequently the phenomenology of radion field has 
also been studied extensively \cite{GW_radion,kribs,julien,wolfe}. This radion phenomenology  
along with RS graviton modes \cite{dhr,rizzo,yong,dhr1,thomas} and RS black holes 
\cite{wiseman,naresh,dai} are considered as key signatures to 
search for warped extra dimension in collider experiments in LHC \cite{atlas1,atlas2}. Some variants of RS model and its modulus stabilization 
have been discussed in \cite{ssg1,csaki,tp1,tp2,tp3}.

However, the important questions that remain are: 
\begin{itemize}
 \item Can the modulus of RS like warped spacetime be stabilized by some other agent 
 than that of a massive bulk scalar field?
 
 \item If such a stabilization mechanism is found, then what is the resulting 
 stabilization condition? Moreover, what are the mass and the coupling (with SM fields) of 
 radion field?
\end{itemize}

We aim to address these questions in this paper and for this purpose, the effective on-brane theory 
we used, is formulated by Kanno-Soda in \cite{kanno}. 

The paper is organized as follows : Following two sections are devoted to brief reviews of RS scenario and 
the description of effective on-brane theory for RS like spacetime. The possibility of modulus stabilization 
for the said geometric model is explored in section IV. 
In section V, we find the necessary constraints on parametric space in order to solve the gauge hierarchy problem. 
Section VI is reserved for the coupling between 
radion and SM fields. We end the paper with some conclusive remarks.

\section{Brief description of RS scenario and its stabilization via GW mechanism}

RS scenario is defined on a five dimensional AdS spacetime involving one warped and compact 
extra spacelike dimension. Two 3-branes known as visible/TeV and hidden/Planck brane 
are embedded in this five dimensional spacetime where the intermediate region 
between the branes is termed as 'bulk'. If $y$ is the extra dimensional linear 
coordinate, then the branes are 
located at two fixed points $y=(0,\pi r_c)$ while the latter one is identified 
with our usual four dimensional universe. Here, $r_c$ is the compactification radius of the extra dimension. 
The opposite brane tensions along with the finely 
tuned five dimensional cosmological constant serve as energy-momentum tensor of RS 
scenario. The resulting spacetime metric \cite{RS1} is non-factorizable and expressed as,
\begin{equation}
ds^2 = e^{- 2 kr_c|\phi|} \eta_{\mu\nu} dx^{\mu} dx^{\nu} -dy^2 \label{eq1}
\end{equation} 
Due to $S^1/Z_2$ compactification along the extra dimension, $y$ ranges from 
$-\pi r_c$ to $+\pi r_c$.
The quantity $k=\sqrt{\frac{-\Lambda}{12M^3}}$, is of the order of 5-dimensional Planck
scale $M$. Thus $k$ relates the 5D Planck scale $M$ to the 5D cosmological constant
$\Lambda$.\\
All the dimensionful parameters described above are
related to the reduced 4-dimensional Planck scale ${M}_{Pl}$ as,
\begin{equation}
 M_{Pl}^2=\frac{M^3}{k}(1-e^{-2k\pi r_c})\label{rplanckmass}
\end{equation}

In order to solve the hierarchy problem, it is assumed in RS scenario that the branes 
are separated by such a distance that 
$k\pi r_c \approx 36$. Then the exponential factor present
in the metric, which is often called warp factor, produces 
a large suppression so that a mass scale of the order of Planck scale is reduced to TeV scale on the 
visible brane. 
A scalar mass say mass of Higgs is given as, 
\begin{equation}
 m_H=m_{0}e^{-k\pi r_c}\label{physmass}
\end{equation}
where $m_H$ and $m_0$ are physical and bare Higgs mass respectively. 
In RS model, it is assumed that the inter-brane separation is of the order of Planck length 
so that the required hierarchy between the branes is generated. 
One of the crucial aspects of this braneworld scenario is to stabilize 
the distance between the branes.
For this purpose, Goldberger and Wise demonstrated that the modulus corresponding to the
 radius of the extra dimension in RS warped geometry model can be stabilized \cite{GW1} by invoking a massive 
scalar field in the bulk with appropriate vacuum expectation values
(vev) at the two 3-branes that reside at the orbifold fixed points. 
Consequently the phenomenology of the radion 
field originating from 5D gravitational degrees of freedom has also been explored \cite{GW_radion}.

\section{Low energy expansion scheme and effective four dimensional action for RS like spacetime}
In RS model, the Einstein equations are derived for a fixed inter-brane separation as 
well as for flat 3-branes. However, the scenario changes if the distance between the branes 
becomes a function of spacetime coordinates and the brane geometry is curved. These 
generalizations are incorporated while deriving the effective on-brane action via 
``low energy expansion scheme'' in \cite{kanno}.

The model we considered in the present paper is described by a five dimensional 
anti-de Sitter (AdS) spacetime with two 3-branes embedded within the spacetime. 
The spacetime geometry is $S^1/Z_2$ orbifolded along the extra dimension. 
Taking $y$ as the extra dimensional 
linear coordinate, the branes are situated at orbifolded fixed points i.e. at $y=0$ 
(Planck brane) and $y=l$ (TeV brane) 
respectively. An additional constant energy density (over and above the brane tension) is localized 
on Planck brane. 
Moreover, the proper distance between the branes is considered as a function of spacetime coordinates. 
The action of this model \cite{kanno} is following: 
\begin{eqnarray}
 S&=& \frac{1}{2\kappa^2} \int d^4x dy \sqrt{-G} [R^{(5)}+ (12/l^2)]\nonumber\\ 
 &-&\int d^4x [\sqrt{-g_{hid}} V_{hid} + \sqrt{-g_{vis}} V_{vis}]\nonumber\\
 &-&\int d^4x \sqrt{-g_{hid}} \Lambda_{hid}
 \label{five dim action}
\end{eqnarray}
with $x^\mu=(x^0,x^1,x^2,x^3)$ are the brane coordinates. 
$\frac{1}{2\kappa^2} = M^3$, $M$ is the five dimensional Planck mass. $R^{(5)}$ and 
$l$ ($\sim$ Planck length) are the Ricci scalar and curvature radius of the five dimensional spacetime. 
Hidden and visible brane tensions are respectively, given by, $V_{hid} = \frac{6}{\kappa^2l}$ and 
$V_{vis} = -\frac{6}{\kappa^2l}$. $\Lambda_{hid}$ is the 
additional energy density, localized on the hidden brane.

We use the following metric ansatz \cite{kanno} with a warping nature along the extra dimension i.e. 
\begin{equation}
 ds^2 = e^{2\phi(x)}dy^2 + e^{-2A(y,x)}h_{\mu\nu}(x)dx^\mu dx^\nu
 \label{five dim metric}
\end{equation}
$A(y,x)$ is commonly known as warp factor which has a dependence  
on all the spacetime coordinates while the component $G_{yy}$ 
is taken as the function of brane coordinates only \cite{GW_radion}. $\Lambda_{hid}$ is assumed to be less 
than the brane tensions. As a consequence of this assumption, brane curvature radius $L$ is much 
larger than the bulk curvature $l$ i.e. $\epsilon= (\frac{l}{L})^2 \ll 1$. Then the bulk Einstein equations 
can be solved perturbatively where $\epsilon$ is taken as the perturbation parameter. This method 
is known as ``low energy expansion scheme'' \cite{kanno} in which the metric is expanded with increasing 
power of $\epsilon$. The zeroth order perturbation solution replicates the RS solution 
where the inter-brane separation is constant. The effective on-brane action obtained up to 
first order of perturbation, incorporates the fluctuation of modulus as well as non-zero value 
of brane matter. Using the low energy expansion scheme, the warp factor and the effective four dimensional 
action are as follows:

\begin{equation}
 A(y,x) = \frac{y}{l}e^{\phi(x)}
 \label{warp factor}
\end{equation}
and
\begin{eqnarray}
 S_{eff}&=&\frac{l}{2\kappa^2} \int d^4x \sqrt{-h} [\Psi(x)R^{(4)} 
 - \frac{3}{2(1-\Psi)}h^{\mu\nu}\partial_\mu\Psi\partial_\nu\Psi]\nonumber\\
 &-&\int d^4x \sqrt{-h} \Lambda_{hid}
 \label{effective action}
\end{eqnarray}

where $\Psi(x) = [1-\exp{(-2e^\phi(x))}]$ and $R^{(4)}$ is the Ricci scalar formed by $h_{\mu\nu}$.
This above form of $S_{eff}$ matches with that obtained in \cite{GW_radion}.

It may be noticed from eqn.(\ref{effective action}) that upon 
projecting the bulk gravity on the brane, the extra degrees of freedom of $R^{(5)}$ 
(with respect to $R^{(4)}$) appears as scalar field $\Psi(x)$ which directly couples with 
the four dimensional Ricci scalar. Hence the effective on-brane action 
is a Brans-Dicke like theory where the self coupling of the scalar field is not a constant.

Eqn.(\ref{five dim metric}) leads to the separation between hidden and visible brane along the path 
of constant $x^\mu$ as follows :
\begin{equation}
 d(x) = \int_{0}^{l} dy e^{\phi(x)} = l e^{\phi(x)}
 \label{brane separation}
\end{equation}
Above expression (eqn.(\ref{brane separation})) clearly indicates that the proper distance 
between the branes depends on the brane coordinates and thats why 
$d(x)$ can be treated as field. From the perspective of four dimensional effective theory, 
this field is termed as 'radion field' (or modulus field) which is symbolized by $\Psi(x)$ in eqn.(\ref{effective action}). 
Designating the modulus field, now we move to the issue for stabilizing that modulus which is described in the next section.

\section{Modulus stabilization}
Like other higher dimensional braneworld scenario, one of the important aspects of the model we considered, 
is to stabilize the 
modulus. For this purpose, one needs to generate a suitable radion potential which has a stable minimum. 
Goldberger and Wise proposed a mechanism \cite{GW1,GW_radion} to create such a radion potential by invoking a massive bulk scalar 
field with appropriate boundary values. 

In this paper, we show that without the presence of a massive scalar 
field in the bulk \cite{GW1}, a non-zero constant energy density localized on hidden brane ($\Lambda_{hid}$) 
is sufficient to generate a stable radion potential. In the next subsection, we explicitly describe the mechanism 
for generating the modulus potential from ``hidden brane energy density'' ($\Lambda_{hid}$).

\subsection{Generating the radion potential from brane localized term}
The form of effective action obtained in eqn.(\ref{effective action}) clearly reveals that the kinetic 
term of the gravitational field ($h_{\mu\nu}(x)$) as well as of the radion field ($\Psi(x)$) are not 
canonical. Thus to make the kinetic terms canonically normalized, lets transform the 
gravitational field as,
\begin{equation}
 h_{\mu\nu}(x) \longrightarrow \tilde{h}_{\mu\nu}(x) = e^{-\sigma(\Psi)}h_{\mu\nu}(x)
 \label{conformal transformation}
\end{equation}
The above transformation is commonly known as conformal transformation where $\sigma(\Psi)$ is 
the conformal factor. Due to this conformal transformation, the Ricci scalar tranforms as follows,

\begin{equation}
 R^{(4)}(h) = e^{-\sigma} [\tilde{R^{(4)}}(\tilde{h}) - 3\tilde{\square}\sigma - \frac{3}{2}\tilde{h}^{\mu\nu}\partial_\mu\sigma\partial_\nu\sigma]
 \nonumber\\
\end{equation}

where $R^{(4)}(h)$ and $\tilde{R}^{(4)}(\tilde{h})$ are the Ricci scalar formed by $h_{\mu\nu}$ and $\tilde{h}_{\mu\nu}$ 
respectively. The d'alembertian operator is defined as $\tilde{\square} = \tilde{h}^{\mu\nu}\tilde{\nabla}_\mu\partial_\nu$. 
Using the above relaton between $R$ and $\tilde{R}$, the action (in eqn.(\ref{efective action})) can be written as,

\begin{eqnarray}
 S_{eff}&=&\frac{l}{2\kappa^2} \int d^4x \sqrt{-\tilde{h}} [\Psi e^{\sigma}(\tilde{R}^{(4)} 
 - 3\tilde{\square}\sigma - \frac{3}{2}\tilde{h}^{\mu\nu}\partial_\mu\sigma\partial_\nu\sigma)\nonumber\\
 &-&\frac{3}{2(1-\Psi)}e^{\sigma}\tilde{h}^{\mu\nu}\partial_\mu\Psi\partial_\nu\Psi]
 - \int d^4x \sqrt{-\tilde{h}} e^{2\sigma}\Lambda_{hid}
 \nonumber
\end{eqnarray}

In order to make the redefined gravitational field ($\tilde{h}_{\mu\nu}$) canonical, take the conformal factor as 
$e^{\sigma(\Psi)} = \frac{1}{\Psi}$. With this choice of conformal factor, the above action turns out to be,

\begin{eqnarray}
 S_{eff}&=&\frac{l}{2\kappa^2} \int d^4x \sqrt{-\tilde{h}} [\tilde{R}^{(4)} 
 - \frac{3}{2}\frac{1}{\Psi^2(1-\Psi)}\tilde{h}^{\mu\nu}\partial_\mu\Psi\partial_\nu\Psi]\nonumber\\
 &-&\int d^4x \sqrt{-\tilde{h}} \frac{1}{\Psi^2}\Lambda_{hid}
 \label{efective action1}
\end{eqnarray}

where we drop the surface term constructed by the d'alembertian operator. So, we manage to make the gravitational 
field canonical which is evident from the action presented in eqn.(\ref{efective action1}).

Comparing eqn.(\ref{effective action}) and eqn.(\ref{efective action1}), it may be observed that the self coupling 
of the scalar field changes on performing the conformal transformation. It is expected because the conformal 
factor itself depends on the scalar field $\Psi(x)$.

To get canonically normalized radion field lets transform $\Psi(x) \rightarrow \chi(x)$ in such a way that 
the following relation 
\begin{equation}
 [\frac{d\chi}{d\Psi}]^2 = (3l/2\kappa^2) \frac{1}{\Psi^2(1-\Psi)}
 \label{diff equation}
\end{equation}
holds. Thus in terms of $\tilde{h}_{\mu\nu}(x)$ and $\chi(x)$, the four dimensional effective 
action takes the following form,

\begin{eqnarray}
 S_{eff} = \int d^4x \sqrt{-\tilde{h}} &[&\frac{l}{2\kappa^2} \tilde{R}^{(4)} 
 - \frac{1}{2}\tilde{h}^{\mu\nu}\partial_\mu\chi\partial_\nu\chi\nonumber\\
 &-&\frac{1}{\Psi(\chi)^2}\Lambda_{hid}]
 \label{efective action2}
\end{eqnarray}

where all the kinetic terms become normalized. The term $\frac{1}{\Psi(\chi)^2}\Lambda_{hid}$ in 
eqn.(\ref{efective action2}) acts 
as a potential for the canonical radion field $\chi(x)$. Afterwards, we denote this potential term by $V(\chi)$. 
In order to extract the explicit dependence of $V(\chi)$ on $\chi$, one needs the relation between $\Psi$ and $\chi$ 
i.e. $\Psi=\Psi(\chi)$. Solving eqn.(\ref{diff equation}), one obtains the dependence of $\Psi$ 
on $\chi$ as follows :
\begin{equation}
 \frac{1}{\Psi(\chi)} = \frac{D}{4} [e^{\frac{b\chi}{2}} + De^{-\frac{b\chi}{2}}]^2
 \label{relation of psi and chi}
\end{equation}

where $D$ is a dimensionless parameter arises in the process of normalizing the radion field and $b=\sqrt{\frac{2\kappa^2}{3l}}$. 
Above expression of $\Psi=\Psi(\chi)$ immediately leads to the radion potential as follows :
\begin{equation}
 V(\chi) = (\frac{D^2}{16}) \Lambda_{hid}[e^{\frac{b\chi}{2}} + De^{-\frac{b\chi}{2}}]^4
 \label{radion potential}
\end{equation}

The potential $V(\chi)$ goes to zero 
as $\Lambda_{hid}\rightarrow 0$ and thus it is clear that the ``radion field potential'' generates entirely 
due to the presence of non-zero constant energy density ($\Lambda_{hid}$) localized on hidden brane. 

In the 
next subsection, we check whether $V(\chi)$ admits any stability or not.

\subsection{Stability of radion potential: Stabilized modulus}
\begin{figure}[!h]
\begin{center}
 \centering
 \includegraphics[width=3.4in,height=2.5in]{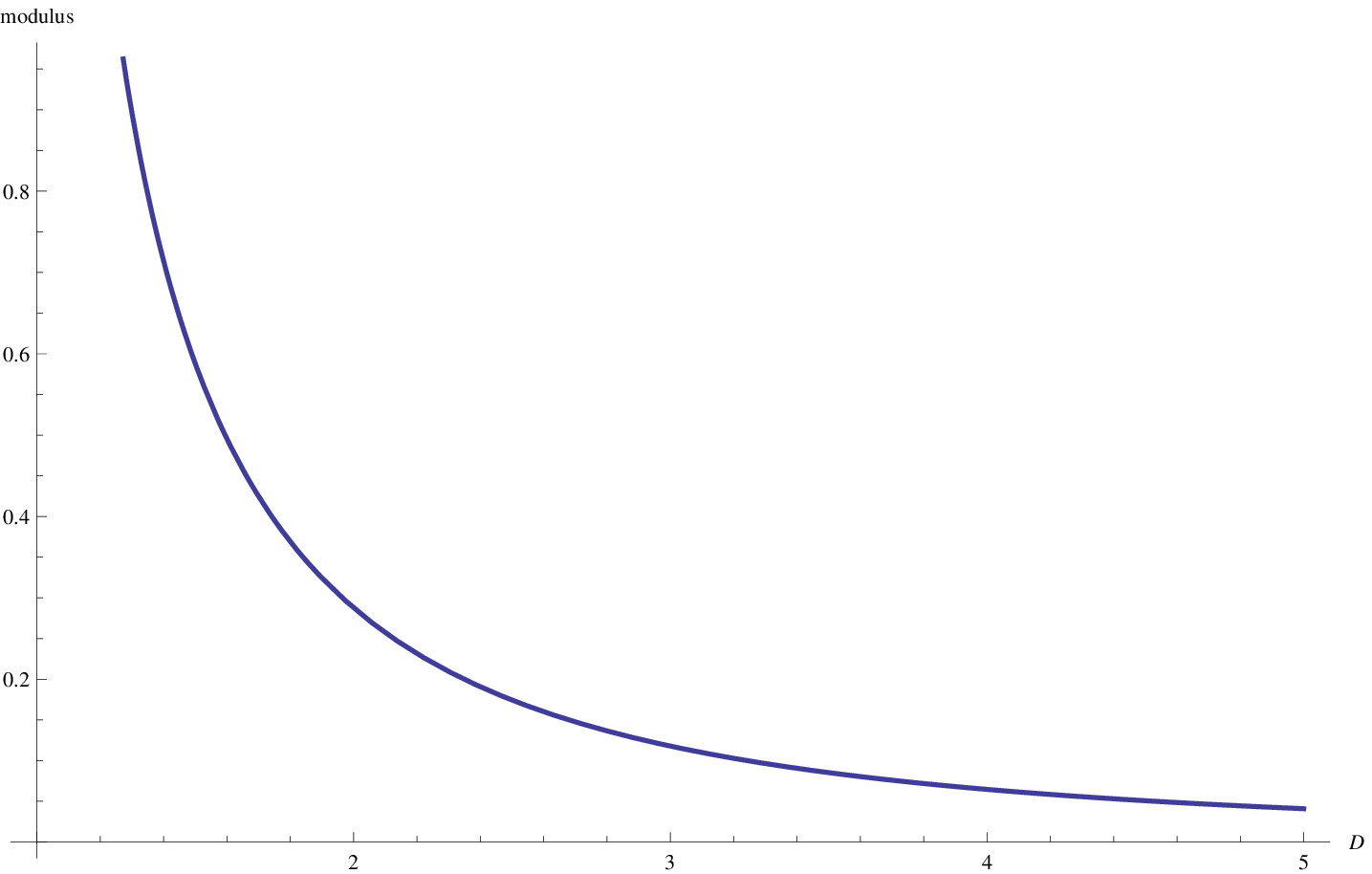}
 \caption{$\frac{<d(x)>}{l}$ vs $D$}
 \label{plot parameter vs brane separation}
\end{center}
\end{figure}

It can be shown that the radion potential $V(\chi)$ has a stable minimum at 
\begin{equation}
  <\chi> = \frac{1}{b} \ln D = \frac{3l}{2\kappa^2} \ln D
  \label{vev of chi}
\end{equation}
if $\Lambda_{hid}$ is considered to be positive. 
Using eqn.(\ref{brane separation}) and eqn.(\ref{relation of psi and chi}), we determine the relation 
between the canonical radion field ($\chi(x)$) and the inter-brane separation ($d(x)$) as follows :
\begin{equation}
 \frac{D}{4} [e^{\frac{b\chi}{2}} + De^{-\frac{b\chi}{2}}]^2 = [1 - e^{-2d(x)/l}]^{-1}
 \label{transformation relation}
\end{equation}
By putting the vacuum expectation value (vev) of $\chi$ (i.e. $<\chi>$) into eqn.(\ref{transformation relation}), we obtain 
the stabilized inter-brane separation ($<d(x)>$) and it is given by,
\begin{equation}
 \frac{<d(x)>}{l} = \frac{1}{2} \ln{[\frac{D^2}{D^2-1}]}
 \label{stabilized modulus}
\end{equation}
The modulus stabilization condition (in eqn.(\ref{stabilized modulus})) clearly reveals that the value of the parameter 
$D$ is constrained to be greater than one otherwise the branes can not be stabilized. Using eqn.(\ref{stabilized modulus}), we 
obtain Figure(\ref{plot parameter vs brane separation}) between $<d(x)/l>$ and $D$.

The figure demonstrates that the modulus decreases with increasing $D$ and the branes are going 
to be collapsed at large value of $D$.

Consequently, we find the mass of radion field excitations about the minimum as follows :
\begin{equation}
 m_{\chi}^2 = \frac{d^2V}{d\chi^2}(<\chi>) = \frac{2\kappa^2}{3l}D^2\Lambda_{hid}
 \label{radion mass}
\end{equation}
$m_{\chi}$ goes 
to zero as $\Lambda_{hid}\rightarrow 0$. This is expected because the radion potential is created due to 
the non-zero value of $\Lambda_{hid}$.

\section{Required conditions for solving gauge hierarchy problem}
Using the form of warp factor (in eqn.(\ref{warp factor})), the physical Higgs mass is obtained as,
\begin{equation}
 m_{phy} = m_0e^{-\frac{<d(x)>}{l}}
 \label{physical mass}
\end{equation}
where $m_0$ is the natural cut-off scale ($10^{19}$ GeV) of the theory 
and to derive the above relation between $m_{phy}$ and $m_0$, it is assumed that 
Higgs field is confined to visible brane only \cite{RS1}. Using eqn.(\ref{stabilized modulus}), $m_{phy}$ can be expressed as,
\begin{equation}
 m_{phy} = m_0\sqrt{\frac{(D^2-1)}{D^2}}
 \label{phy higgs mass}
\end{equation}
Thus in order to confine the Higgs mass ($m_{phy}$) within TeV scale, the value of the parameter $D$ 
is taken as $D= \sqrt{1+10^{-32}}$. 

Moreover from the expression of radion mass in eqn.(\ref{radion mass}), it is clear that 
with $D= \sqrt{1+10^{-32}}$, $m_{\chi}$ comes at TeV scale if the constant 
energy density on hidden brane ($\Lambda_{hid}$) is of the order of $10^{44}$(GeV)$^{4}$.

\section{Coupling of radion with SM fields}
Being a part of gravitational degrees of freedom, radion field interacts 
with brane matter and the interaction lagrangian is  
constrained by four dimensional general covariance. From five dimensional metric 
(in eqn.(\ref{five dim metric})), it is clear that the induced metric on visible brane is 
$\exp{[-A(l,x)]}h_{\mu\nu}$ (where $A(l,x)= \frac{d(x)}{l}$) and consequently $d(x)$ directly couples 
with Standard Model fields confined on visible brane.\\
For example, consider the Higgs sector of Standard Model,
\begin{eqnarray}
 S_{higgs} = \frac{1}{2}\int d^4x \sqrt{-h}&[&e^{-A(l,x)}h^{\mu\nu}\partial_\mu\xi\partial_\nu\xi \nonumber\\
 &+&e^{-2A(l,x)}m_0^2\xi^2]
 \nonumber
\end{eqnarray}
where $\xi(x)$ is Higgs field and recall that $m_0$ is the natural mass scale of the theory. 
In terms of physical Higgs mass (see eqn.(\ref{physical mass})), above action turns out to be, 
\begin{eqnarray}
 S_{higgs}&=&\frac{1}{2}\int d^4x \sqrt{-h}[\exp{(-\frac{d(x)}{l})}h^{\mu\nu}\partial_\mu\xi\partial_\nu\xi \nonumber\\
 &+&\exp{(-2\frac{d(x)}{l})}\exp{(2\frac{<d(x)>}{l})}m_{phy}^2\xi^2]
 \nonumber
\end{eqnarray}

Using the transformation relation between $d(x)$ and $\chi(x)$ (in eqn.(\ref{transformation relation})), 
the interaction lagrangian between Higgs and canonical radion field is given as per following 
expression,
\begin{equation}
 L_{int}[H-\delta\chi] = \delta\chi \sqrt{\frac{2\kappa^2}{3l}}\exp{(\frac{<d(x)>}{l})} T_\mu^\mu[\xi]
 \nonumber
\end{equation}
where $\delta\chi$ is the fluctuation of radion field about its vev (i.e. $\chi=<\chi>+\delta\chi$) and 
$T_\mu^\mu[\xi]$ is the trace of energy-momentum tensor of Higgs field. 
So, the coupling between radion and Higgs field becomes, 
$\lambda(H-\delta\chi) = \sqrt{\frac{2\kappa^2}{3l}}\exp{(\frac{<d(x)>}{l})}m_{phy}^2$. 
Similar consideration holds for other SM fields. For example for Z-boson, 
$\lambda(Z-\delta\chi) = \sqrt{\frac{2\kappa^2}{3l}}\exp{(\frac{<d(x)>}{l})}m_Z^2$, where 
$m_Z$ is mass of Z-boson. 
Thus the stabilized inter-brane separation plays a crucial role in determining the coupling strength 
between radion and SM fields. In the present case, 
\begin{equation}
 \frac{<d(x)>}{l} = \frac{1}{2}\ln[\frac{D^2}{(D^2-1)}]
 \nonumber
\end{equation}
Hence finally we arrive at,
\begin{equation}
 \lambda(H-\delta\chi) = \sqrt{\frac{2\kappa^2}{3l}}\sqrt{\frac{D^2}{(D^2-1)}}m_{phy}^2
 \label{coupling1}
\end{equation}
and
\begin{equation}
 \lambda(Z-\delta\chi) = \sqrt{\frac{2\kappa^2}{3l}}\sqrt{\frac{D^2}{(D^2-1)}}m_Z^2
 \label{coupling2}
\end{equation}
Eqn.(\ref{coupling1}) and eqn(\ref{coupling2}) clearly indicate that for $D= \sqrt{1+10^{-32}}$ 
(considered earlier to solve the hierarchy problem), the coupling between 
radion and SM fields is of the order of TeV scale.\\

\section{Conclusion}
We consider a five dimensional AdS, compactified warped geometry model 
with two 3-branes embedded within the spacetime. An additional constant energy density (over and above the brane tension) 
is localized on hidden brane. In this model, 
we examine the possibility of modulus stabilization. The findings and implications of 
our results are as follows:

\begin{itemize}
 \item We find that the constant energy density ($\Lambda_{hid}$) localized on hidden brane 
 is sufficient to generate a potential term for the radion field. 
 This potential has a stable minimum as long as $\Lambda_{hid}$ is positive. Moreover, the radion potential ($V(\chi)$)
 goes to zero as $\Lambda_{hid}\rightarrow 0$ 
 and thus it is clear that $V(\chi)$ is created due to the presence of ``hidden brane energy density''.
 
 \item The stabilization condition is determined and given in eqn.(\ref{stabilized modulus}). 
 From the perspective of modulus stabilization, possible value of the parameter $D$ which 
 arises on the process of normalization of radion field, is scanned. As a result, 
 the parameter $D$ is constrained to be greater than one otherwise it becomes 
 impossible to stabilize the two brane system.  
 Furthermore, the stabilized value of the modulus decreases with increasing $D$ and consequently  
 the branes are going to be collapsed for large $D$, which is 
 clear from Figure(\ref{plot parameter vs brane separation}).
 
 \item We also find the mass (see eqn.(\ref{radion mass})) and coupling (with SM fields, see 
 eqn.(\ref{coupling1}) and eqn.(\ref{coupling2})) of radion field in this higher dimensional braneworld 
 scenario. It is shown that by considering $D= \sqrt{1+10^{-32}}$ and $\Lambda_{hid}= 10^{44}$(GeV)$^{4}$, 
 mass as well as coupling of radion field can be kept at TeV scale without sacrificing the necessary 
 conditions to solve the gauge hierarchy problem.
 
\end{itemize}

\section*{Acknowledgements}
  I thank S. SenGupta and S. Chakraborty for illuminating discussions.

\end{document}